\documentclass[aps,pra,twocolumn,superscriptaddress]{revtex4}
\usepackage{bbding}
\usepackage{amsthm}
\usepackage{dcolumn}
\usepackage{bm}
\usepackage{subfigure}
\usepackage{epsfig,graphicx,times}
\usepackage{amstext}
\usepackage{amsmath}
\usepackage{amssymb}
\usepackage{graphicx}
\usepackage{latexsym}
\usepackage{bm}
\usepackage[colorlinks,citecolor=blue, linkcolor=blue,hyperindex,dvipdfm]{hyperref}

\begin{document}

\title{Nanomechanically induced transparency in $\mathcal{PT}$-symmetric
optical cavities}
\author{Amjad Sohail}
\email{amjadsohail@gcuf.edu.pk}
\affiliation{Department of Physics, G.C University, Allama Iqbal Road, Faisalabad 38000,
Pakistan}
\author{Rizwan Ahmed}
\affiliation{Physics division, Pakistan institute of nuclear science and technology
(PINSTECH), Nilore, Islamabad 45650 Pakistan}
\author{Hazrat Ali}
\affiliation{Department of Physics, Abbottabad University of Science and Technology, P.O. Box 22500 Havellian KP, Pakistan}

\begin{abstract}
In this paper, we analytically present the phenomena of nanomechanically induced transparency (NMIT) and transmission rate in a parity-time-symmetric ($\mathcal{PT}$-symmetric) opto-nanomechanical system (ONMS) where a  levitated dielectric nanospheres is trapped near the antinodes closest to right mirror of passive cavity which further coupled to an active cavity via hoping factor.
We find that the phenomenon of NMIT may be generated from the output probe field in the presence of an effective opto-nanomechanical coupling between the cavity field and the nanosphere, whose steady-state position is influenced by the Coulomb interaction between the cavity mirror and the nanosphere.
In addition, the width and height of the transparency window can be controlled through the effective optomechanical coupling, which is readily adjusted by altering changing the nanosphere's radius and the Coulomb interaction. One of the most interesting result is the transition NMIT behavior in $\mathcal{PT}$-symmetric and broken $\mathcal{PT}$-symmetric regime. We show that the presence of nanosphere in the passive cavity enhances the width and transmission rate of NMIT window in
passive-passive regime and in passive-active regime, a notable decrease of sideband amplification has been observed. These results show that our scheme may find some potential applications for optical signal processing an and quantum information processing.
\end{abstract}
\maketitle
\section{Introduction}
One basic axiom in canonical quantum mechanics requires that
the Hermiticity of every operator be directly related with a physical
observable. In addition, the spectrum of a Hermitian operator is guaranteed
to be real. Bender and colleagues, however, discovered a class of
Hamiltonians that are non-Hermitian with fully real spectra provided they
fulfil combined parity-time ($\mathcal{PT}$) symmetry \cite{bender1,bender2}%
. Open physical systems with balanced amplification (gain) and absorption (loss) are known as $\mathcal{PT}$-symmetric systems. If a critical value is exceeded for the parameter controlling the degree of non-Hermiticity, such systems exhibit spontaneous symmetry breaking.
The eigenvalues of such system become complex beyond this threshold even though $[PT,H]=0$ \cite{BPENG}. $\mathcal{PT}$%
-symmetry has been widely investigated theoretically and experimentally
demonstrated in various physical systems, with quantum optics emerging as
the most versatile platform to explore $\mathcal{PT}$-symmetric
applications. Up to now, $\mathcal{PT}$-symmetric systems have been
extensively observed in quantum information processing (QIP) and quantum
optics, opening up novel applications including soliton active controlling
\cite{RDBA,FNAZ} enhancing photon blockade \cite{JLRY}, realizing quantum
chaos \cite{XYLU}, strengthening optical nonlinearity \cite{JLXZ,SKGA} and
so on. In addition, $\mathcal{PT}$-symmetric Hamiltonian has also been
experimentally realized in many physical systems \cite{BPNG,AREG,LFENG,LCHAN}.

Although $\mathcal{PT}$-symmetry in optomechanical system enhances various
quantum optical effects when balanced through the loss with extra gain,
there still are opposite directions in which environmental loss play a vital
role. As an analog of electromagnetically induced transparency (EIT), optomechanically induced transparency (OMIT) is the most notable of these directions \cite{Agr,SAZY,SAUM}. It is notable that the value of the
cavity decay rate must be large in order to observe
optomechanically-induced-transparency (OMIT). It is also worth mentioning
that an extra gain in optomechanical system is taken into account to deviate
from OMIT behavior, even leads to optomechanically-induced absorption (OMIA)
or inverted-OMIT. As of right now, other additional coupled systems such as two and three-level atoms \cite{SAZY,YXYF,YTYL,YHJC,YCTS}, Bose-Einstein condensates \cite{BFRS,RITS,YKAA}, magnon \cite{S1,S2,S3,S4} etc., are still needed for the enhancing quantum properties like OMIT, entanglement, 
Furthermore, levitated nanosphere, trapped inside an
optical cavity, has also been considered to discuss many quantum features
such as gravitational wave detection \cite{KRHG,AAGA} and quantum ground
state cooling \cite{KNBF,MJFP,CDER,RIOP}. Therefore, a natural question
arises. Can an optically trapped nanosphere play an important role in $%
\mathcal{PT}$-symmetric coupled cavity system?

In this paper, we address this question by considering $\mathcal{PT}$%
-symmetric system that comprise two coupled cavities. The passive cavity contain a levitated nanosphere. Particularly, we show: (i) existence of an NMIT window in the red-sideband region and a small dip in the blue-sideband region in a single cavity system; (ii) NMIT-like spectrum in addition to sharp absorptive peak in the blue-sideband region when we consider an active cavity with no loss/gain; (iii) an inverted NMIT which strongly depends on the modulated interaction between the cavity and the nanosphere; and (iv) a reversed opto-nanomechanical coupling dependence of the transmission rate.

The rest of the manuscript is organized as follows. In section II, the model and
the corresponding system dynamics of the opto-nanomechanical system are presented. In section III,
we theoretically discuss the NMIT and the transmission of the ONMS. The
section IV concludes our paper.
\begin{figure}[b!]
\centering
\includegraphics[width=1\columnwidth,height=1in]{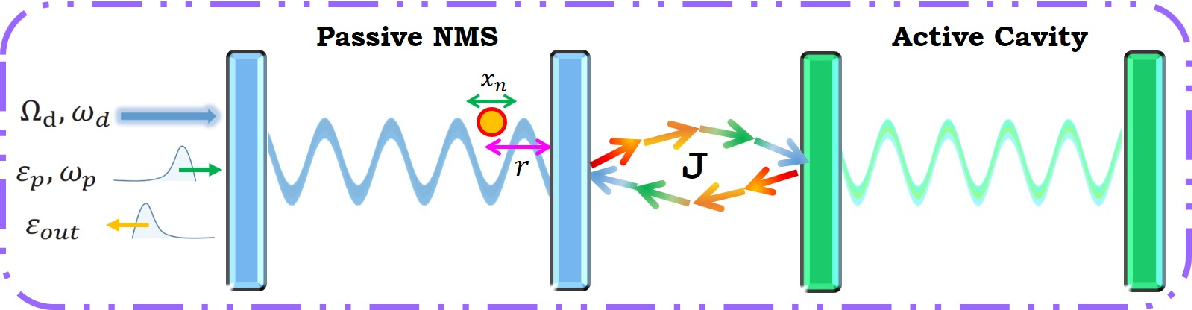} \centering
\caption{(a)\textbf{NMIT in active-passive-coupled cavities,
with a tunable gain-loss ratio}. The passive cavity, which is coupled to an
active cavity through hoping factor, is probed by an external probe and pump
fields. The nanosphere is coupled to the right fixed cavity mirror of the passive
cavity via coulomb interaction.}
\end{figure}
\section{The model and the dynamics}
The current opto-nanomechanical system is comprised of two connected cavities, as depicted in Figure 1, where
the first (passive) cavity contains a nanosphere. The cavity bosonic creation and annihilation operators are represented as $c^{\dagger}_{1}$ and $c_{1})$ while cavity decay rate is $\gamma$.
The second cavity is an active cavity with gain
rate $\kappa$ and is represented by the bosonic creation (annihilation) as $%
c^{\dagger}_{2}(c_{2})$. A coupling field (probe field) with amplitude $\Omega _{d}=\sqrt{2\gamma \wp_{d} /\hbar \omega _{d}}$ ($\varepsilon _{p}=\sqrt{2\gamma \wp_{p} /\hbar \omega _{p}}$), where $\wp_{d}$ ($\wp_{p}$) is the power, $\omega _{d}$ ($\omega _{p}$) is the power frequency of coupling (probe) field and $\gamma$ is the decay rate of the cavity field, drives the passive cavity.
Furthermore, the nanosphere is supposed to be
optically levitated inside the cavity. The driven single mode cavity along
with Coulomb interaction is responsible for the levitation of nanosphere.
The Coulomb interaction (coupling strength) can be realized by direct interaction between the charged on the nanosphere and the left cavity mirror and \cite{HWKU,NWCA,SRCT}. Coupling between the charged left cavity mirror and nanosphere can achieve the Coulomb interaction.
This leads to a effective nonzero coupling between the
nanosphere and the cavity \cite{wenj}.

The total Hamiltonian of the system describing the optonanocal system can be
expressed as
\begin{equation}
H_{T}=H_{0}+H_{I}+H_{d}
\end{equation}%
\begin{equation}
H_{0}=\hbar \omega _{c}\left( c_{1}^{\dagger }c_{1}+c_{2}^{\dagger
}c_{2}\right) +\frac{p_{n}^{2}}{2m},
\end{equation}%
\begin{eqnarray}
H_{I} &=&-\hbar g_{n}c^{\dagger }c\cos ^{2}\left( kx_{n}\right) +\hbar
J(c_{1}^{\dagger }c_{2}+c_{2}^{\dagger }c_{1})  \notag \\
&&-\frac{Q_{1}Q_{2}}{4\pi \varepsilon _{\circ }\left( r-x_{n}\right) },
\end{eqnarray}%
\begin{eqnarray}
H_{d} &=&i\hbar \Omega _{d}\left( c_{1}^{\dagger }e^{-i\omega
_{d}t}-c_{1}e^{-i\omega _{d}t}\right)  \notag \\
&&+i\hbar \left( \varepsilon _{p}c_{1}^{\dagger }e^{-i\omega _{p}t}-\varepsilon
_{p}^{\ast }c_{1}e^{-i\omega _{p}t}\right).
\end{eqnarray}%
The two terms in $H_{0}$ represent the free Hamiltonian of the cavity fields
and the nanosphere. $H_{I}$ contains three interaction terms where the first
term represents the interaction between the cavity field and the nanosphere
with opto-nanomechanical coupling $g_{n}=\frac{3V}{4V_{c}}\frac{\varepsilon _{\circ
}-1}{\varepsilon _{\circ }+2}\omega _{c}$, second term represents the
interaction between the between the two cavity fields with hoping factor $J$
and the last term denotes the Coulomb interaction between the
nanosphere with charge $Q_{1}$ and the cavity mirror with charge $Q_{2}$
\cite{HWKU,NWCA,SRCT,SRCTF}. The two terms in $H_{d}$ represent the cavity
field driven by the controlled (probe) field with amplitude $\Omega_{d}$ ($%
\varepsilon_{p}$) and frequency $\omega_{d}$ ($\omega_{p}$) .

In the rotating frame at the coupling frequency $\omega _{d}$,
\begin{eqnarray}
H &=&\hbar \Delta _{d}\left( c_{1}^{\dagger }c_{1}+c_{2}^{\dagger
}c_{2}\right) +\frac{p_{n}^{2}}{2m} -\hbar g_{n}c_{1}^{\dagger }c_{1}\cos ^{2}\left( kx_{n}\right)  \notag \\
&&+\hbar
J(c_{1}^{\dagger }c_{2}+c_{2}^{\dagger }c_{1}) +i\hbar \Omega _{d}\left( c_{1}^{\dagger }-c_{1}\right)-\sigma x_{n}   \notag \\
&&+i\hbar \left(
\varepsilon _{p}c_{1}^{\dagger }e^{-i\Delta t}-\varepsilon _{p}^{\ast
}c_{1}e^{-i\Delta t}\right),
\end{eqnarray}%
where $\Delta _{d}=\omega _{c}-\omega _{d}$ and $\Delta =\omega _{p}-\omega
_{d}$. We make the assumption that the nanosphere's oscillations are substantially lower than $x_{n}\ll r$, the distance between the levitated nanosphere and the left cavity mirror. As a result, we assumed the Coulomb interaction upto first order approximation, such that $\sigma =\frac{%
Q_{1}Q_{2}}{4\pi \varepsilon _{\circ }r^{2}}$.

The nonlinear Langevin equations that governs the evolution of the operators in this system can be expressed as
\begin{eqnarray}
\dot{x}_{n} &=&p_{n}/m,  \notag \\
\dot{p}_{n} &=&-\hbar g_{n}kc_{1}^{\dagger }c_{1}\sin \left( 2kx_{n}\right) +\sigma
-\gamma _{n}p_{n},  \notag \\
\dot{c}_{1} &=&-\left( i\Delta _{d}+\gamma \right) c_{1}+iJc_{2}+ig_{n}\cos
^{2}\left( kx_{n}\right) +\Omega _{d}  \notag \\
&&+\varepsilon _{p}e^{-i\Delta t},  \notag \\
\dot{c}_{2} &=&-\left( i\Delta _{d}-\kappa \right) c_{2}+iJc_{1}
\label{EQN6}
\end{eqnarray}%
where $\gamma$, $\kappa $ and$\gamma _{n}$ denote the leakage of passive
cavity, leakage of active cavity and damping rates of the nanosphere,
respectively.

Now, we make the ansatz to achieve the steady-state solutions for the strong driving $\Omega _{d}$ and to the lowest order in the weak probe $\varepsilon_{p}$, .
\begin{equation}
\left(
\begin{array}{c}
x \\
c_{1} \\
c_{2}%
\end{array}%
\right) =\left(
\begin{array}{c}
x_{s} \\
c_{1s} \\
c_{2s}%
\end{array}%
\right) +\left(
\begin{array}{c}
x^{+} \\
c_{1}^{+}{} \\
c_{2}^{+}%
\end{array}%
\right) e^{-i\Delta t}+\left(
\begin{array}{c}
x^{-} \\
c_{1}^{-} \\
c_{2}^{-}%
\end{array}%
\right) e^{i\Delta t},  \label{OP}
\end{equation}%
which leads to the steady-state values of the operators as

\begin{eqnarray}
p_{ns} &=&0,x_{ns}=\frac{1}{2k}\sin ^{-1}\left( \frac{\sigma }{\hbar
g_{n}\left\vert c_{1s}\right\vert ^{2}}\right) ,  \notag \\
c_{2s} &=&\frac{-iJ\Omega _{d}}{\left( \gamma +i\Delta _{c}\right) \left(
\kappa -i\Delta _{d}\right) +J^{2}},  \notag \\
c_{1s} &=&\frac{\left( \kappa -i\Delta _{d}\right) \Omega _{d}}{\left(
\gamma +i\Delta _{c}\right) \left( \kappa -i\Delta _{d}\right) +J^{2}},
\end{eqnarray}%
where $\Delta _{c}=\Delta _{d}-g_{n}\cos ^{2}\left( kx_{ns}\right) $.
One may obtain the following quantum Langevin equations for the fluctuation:
\begin{eqnarray}
\delta \ddot{x}_{n} &=&\gamma _{n}\delta \dot{x}-\omega _{n}^{2}\delta x_{n}-%
\frac{\hbar }{m}\left( G\delta c^{\dagger }+G^{\ast }\delta c\right)   \notag
\\
\delta \dot{c}_{1} &=&-\left( i\Delta _{c}+\gamma \right) \delta
c_{1}+iJ\delta c_{2}-iG\delta x_{n}+\varepsilon _{p}e^{-i\Delta t},  \notag
\\
\delta \dot{c}_{2} &=&-\left( i\Delta _{d}-\kappa \right) \delta
c_{2}+iJ\delta c_{1}
\end{eqnarray}%
where $G=G_{n}c_{1s}$ is the effective opto-nanomechanical coupling with $%
G_{n}=g_{n}k\sin \left( 2kx_{ns}\right) $ as opto-nanomechanical coupling
coefficient. The effective frequency of the levitated nanosphere
associated its centre of mass oscillation,
$\omega _{n}=\left[ 2\hbar k^{2}\left\vert c_{1s}\right\vert
^{2}\cos \left( 2kx_{ns}\right) /m\right] ^{1/2}$, can be modulated by the steady-state value of the cavity field.
It can be observed that the effective coupling $G_{n}$ and the frequency of the nanosphere $\omega _{n}$ are tunable due to the fact both rely on the steady-state value of the nanosphere.
In addition, it is also very significant to
mention that for a low value of $2kx_{n}$, we obtain a very low effective
opto-nanomechanical coupling coefficient and a high effective frequency of the
nanosphere and vice versa.

Now, we substitute the ansatz $\delta O=O_{-}e^{-i\Delta t}+O_{+}e^{i\Delta t}$ to obtain the response of the cavity field:%
\begin{equation}
c_{1}^{+}=\frac{[\mathcal{G}_{3}\Omega _{2}\Gamma _{n}+\xi \mathcal{G}_{3}%
\mathcal{G}_{4}]\varepsilon _{p}}{\Omega _{1}\Omega _{2}\Gamma _{n}-\xi
\lbrack \Omega _{3}+\mathcal{G}_{3}\mathcal{G}_{4}(\mathcal{G}_{2}-\mathcal{G%
}_{1})]},  \label{CN}
\end{equation}%
where%
\begin{eqnarray}
\Omega _{1} &=&J^{2}+\mathcal{G}_{1}\mathcal{G}_{3},  \notag \\
\Omega _{2} &=&J^{2}+\mathcal{G}_{2}\mathcal{G}_{4},  \notag \\
\Omega _{3} &=&J^{2}(\mathcal{G}_{3}-\mathcal{G}_{4}),  \notag \\
\mathcal{G}_{1} &=&\kappa +i[\Delta _{d}-g_{n}\cos ^{2}\left( kx_{ns}\right)
-\Delta ],\   \notag \\
\mathcal{G}_{2} &=&\kappa -i[\Delta _{d}-g_{n}\cos ^{2}\left( kx_{ns}\right)
+\Delta ],  \notag \\
\mathcal{G}_{3} &=&-\gamma +i(\Delta _{d}-\Delta ),  \notag \\
\mathcal{G}_{4} &=&-\gamma -i(\Delta _{d}+\Delta ),  \notag \\
\Gamma _{n} &=&\omega _{n}^{2}-i\Delta \gamma _{n}-\Delta ^{2},  \notag \\
\xi &=&i\hbar \left\vert G\right\vert ^{2}, \ \ \ \left\vert G\right\vert
^{2}=G_{n}^{2}\left\vert c_{1s}\right\vert ^{2}
\end{eqnarray}%
We may examine the system's response to the probing frequency by looking at the output field, which will help us to theoretically investigate the phenomena of NMIT. In accordance with input-output theory \cite{DFW}:
\begin{equation}
\varepsilon _{out}(t)+\varepsilon _{p}e^{-i\Delta t}+\varepsilon
_{L}=2\gamma c_{1}.  \label{7}
\end{equation}%
In analogy with Eq. (\ref{OP}), we expand the output field to the first
order as

\begin{equation}
\varepsilon _{out}(t)=\varepsilon _{out}^{0}+\varepsilon
_{out}^{+}\varepsilon _{p}e^{-i\Delta t}+\varepsilon _{out}^{-}\varepsilon
_{p}^{\ast }e^{i\Delta t},  \label{INOUT}
\end{equation}%
where $\varepsilon _{0}$, $\varepsilon _{out}^{+}$ and $\varepsilon
_{out}^{-}$ are the components of the output field oscillating at
frequencies $\omega _{L}$, $\omega _{p}$ and $2\omega _{p}-\omega _{L}$.
Thus one can find the following relations by substituting Eq.~(\ref{INOUT})
into Eq.~(\ref{7})%
\begin{eqnarray}
\varepsilon _{out}^{0} &=&2\gamma c_{1}^{0}-\varepsilon _{L},  \notag \\
\varepsilon _{out}^{+} &=&2\gamma c_{1}^{+}-1,  \notag \\
\varepsilon _{out}^{-} &=&2\gamma c_{1}^{-}.
\end{eqnarray}%
Next, we define
\begin{equation}
\varepsilon _{T}=\varepsilon _{out}^{+}+1=2\gamma c_{1}^{+}.  \label{OT}
\end{equation}%
The real part of $\varepsilon _{T}$, is given by $\chi =$Re$\left[
\varepsilon _{T}\right] =2\gamma (c_{1}^{+}+c_{1}^{+\ast })$ which
corresponds to absorption of the output field at the probe frequency
respectively and can be measured via homodyne technique \cite{DFW}.
Therefore, the transmission rate can be acquired as \cite{Jing,ARJT}:%
\begin{equation}
\eta=\left\vert 1-(2\gamma /\varepsilon _{p})c_{1}^{+}\right\vert ^{2}
\end{equation}%
\begin{figure}[tbp]
\centering
\includegraphics[width=1\columnwidth,height=2.5in]{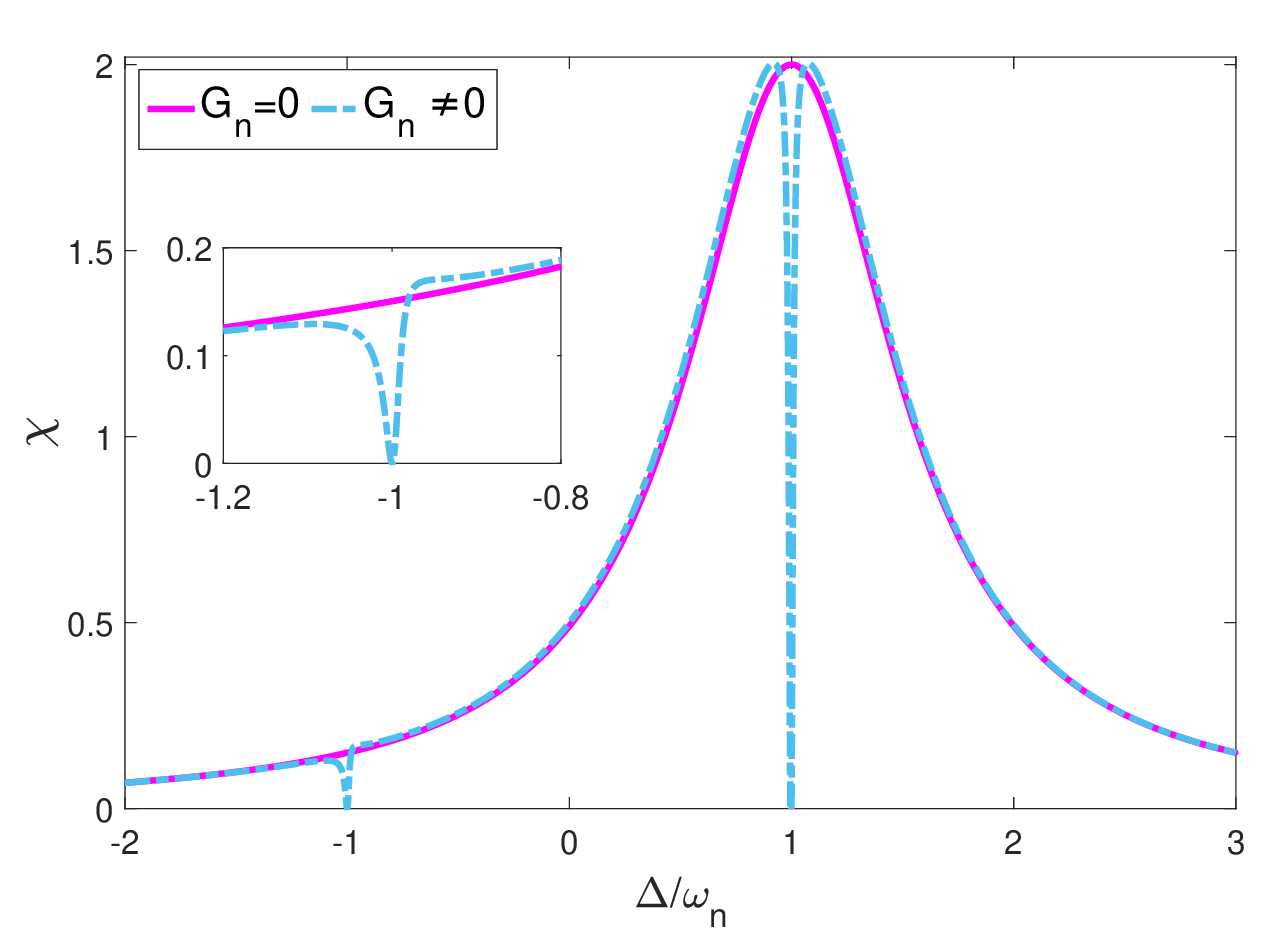} \centering
\caption{The absorption of the output field as a function of $%
\Delta /\protect\omega _{n}$. The others parameters are given by $\protect%
\kappa =2\protect\pi 6\times 10^{4}$Hz, $\wp =0.2$mW, $\protect\lambda =1064$%
nm, $\protect\gamma _{n}=0.003$Hz, sin$(2kx_{n})=0.1$ and $J=0$. The density
of nanosphere $\protect\rho =2300$kg/$m^{3}$, $r=60$nm and the dielectric
constant $\protect\epsilon =2$. The mode volume $V_{c}=(\protect\pi /4)L%
\protect\omega ^{2}$ with mode waist $\protect\omega =20\protect\mu $m and $%
L=0.01$m}
\end{figure}
\section{NMIT in opto-nanomechanical system}
In this section, we numerically demonstrate the optical response of the
system containing a single levitated nanosphere. The parameters which are
same as given in \cite{SGKH,wenj}. The decay rate of the cavity field $%
\kappa =2\pi \times 215$ kHz. The wavelength and the power of the input
laser field are $\lambda =1064$ nm and $\wp=0.2$ mW. In addition, we have
taken a very low damping rate of the levitated nanosphere which is $\gamma
_{n}=0.003$ Hz. To compute the coupling between the levitated nanosphere and the cavity mode, $g_{n}=\frac{3V}{4V_{c}}\frac{\varepsilon
_{\circ }-1}{\varepsilon _{\circ }+2}\omega _{c}$, we consider the radius of the
nanosphere $r=60$ nm, the dielectric constant $\varepsilon _{\circ }=2$, and
the density of the nano oscillator is assumed to be $\rho =2300kg/m^{3}$. In
order to calculate mode volume $V_{c}=\frac{\pi }{4}Lw^{2} $, we have taken
the mode waist $w=20\mu m$ and the length of the cavity $L=25$mm. In addition, we define the probe frequency detuning $\Delta_{p}=\omega _{p}-\omega _{c}$, however, for $\omega_{d}=\omega_{n}$, the probe frequency detuning become $\Delta_{p}=\Delta-\omega_{n}$.

We first demonstrate the optical response of the ONMS by setting $J=0$ i.e., $%
\mathcal{PT}$-symmetric ONMS returns to a typical passive system having decay
rate $\gamma_{n}$ and $\gamma$. In Fig. 2, we exhibit the absorption as a
function of $\Delta/\omega_{n}$ with and without tunable effective
opto-nanomechanical coupling strength. When $x_{ns}=0$, sin($2kx_{ns})$ become zero
and consequently, the effective opto-nanomechanical coupling strength $G_{n}=0$,
which leads to a lorentzian curve around $\Delta/\omega_{n}=1$ as shown in
Fig. 2. Thus, opto-nanomechanical coupling $g_{n}$ between the nanosphere and the
cavity field only trap the nanosphere and don't contribute any additional
interference channel. However, when sin($2kx_{n})=0.1$, a nonzero effective
opto-nanomechanical coupling strength appears which leads to an NMIT window as shown
by the dot-dashed line in Fig. 2. Specifically, there exist two dips which
appears around $\Delta/\omega_{n}=\pm1$ in the presence of $G_{n}$, at the
probe frequency. This is due to the fact that in ONMS, the  nanosphere's damping rate $\gamma_{n}$ is notably small (see the Eq. 11) so that the obtained amplitude
of the output spectra has the minimal values when $\omega_{n}^{2}-\Delta^{2}%
\simeq0$ (i.e., $\Delta=\pm\omega_{n}$), which corresponds to strongest
opto-nanomechanical coupling and ONMS system, in this situation, exhibits the normal
mode splitting. Hence, effective opto-nanomechanical coupling strength gives rise to
symmetric structure, quantified by a NMIT window and two sideband absorption
peaks around $\Delta=\pm\omega_{n}$.

Another fascinating observation is that when we consider the active cavity
has neither loss nor gain. To observe the influence of such active cavity,
with no loss/gain, on the absorption, we have plotted the absorption
spectrum as a function of $\Delta$ for different hoping factor in fig. 3.
Interestingly, in this situation, we not only obtained NMIT around $%
\Delta=\omega_{n}$ but also an additional absorption profile appears around $%
\Delta=-\omega_{n}$. In this case, ONMS still behaves as a passive system
because one cavity has neither loss nor gain and the other is lossy.

\begin{figure}[tbp]
\centering
\includegraphics[width=1\columnwidth,height=2.5in]{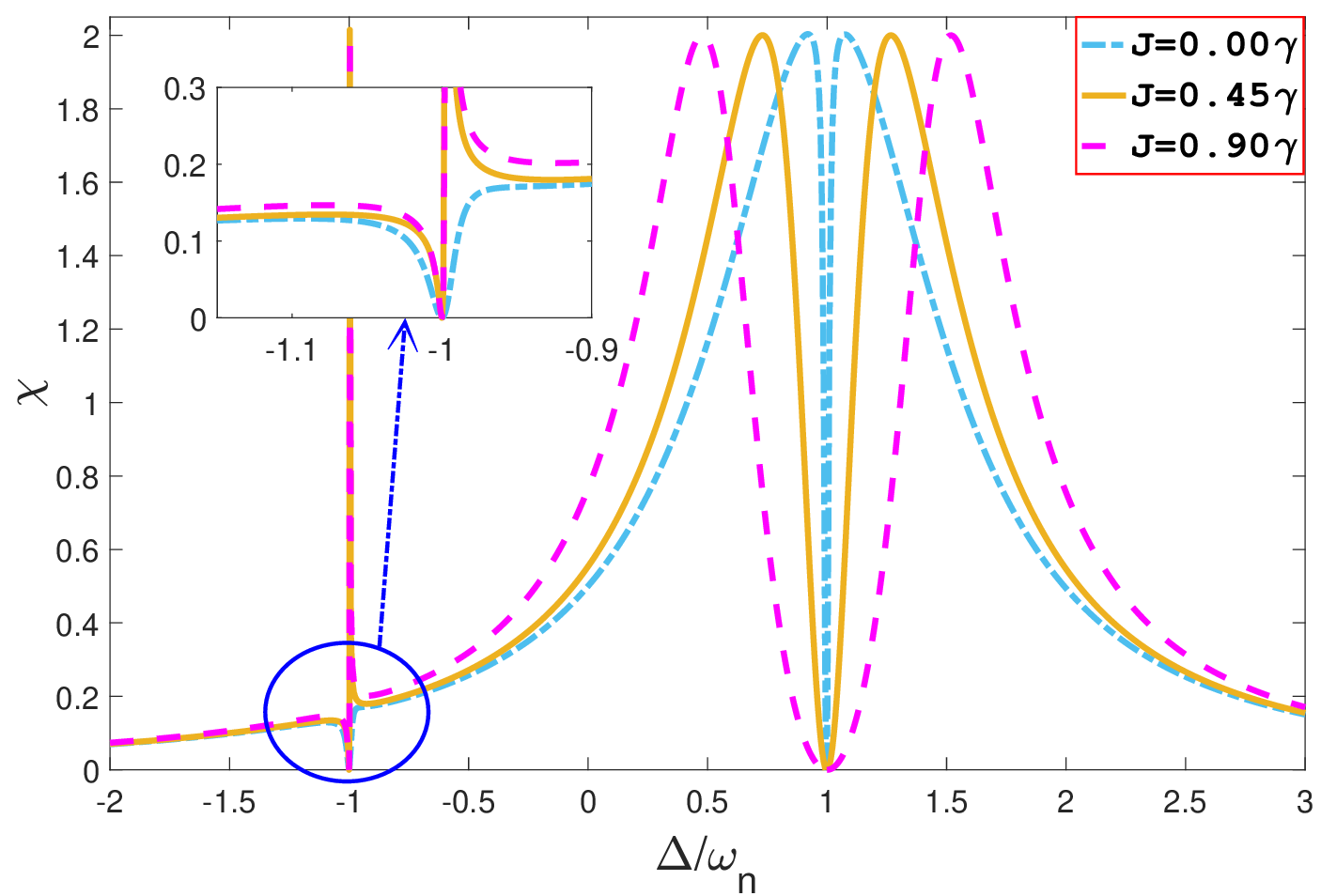} \centering
\caption{The absorption of the output field as a function of
$\Delta /\protect\omega _{n}$ for different value of the hoping factor, with
$\protect\kappa =0$. The others parameters are the same as in Fig. 2.}
\end{figure}
\begin{figure}[tbp]
\includegraphics[width=1\columnwidth,height=2.5in]{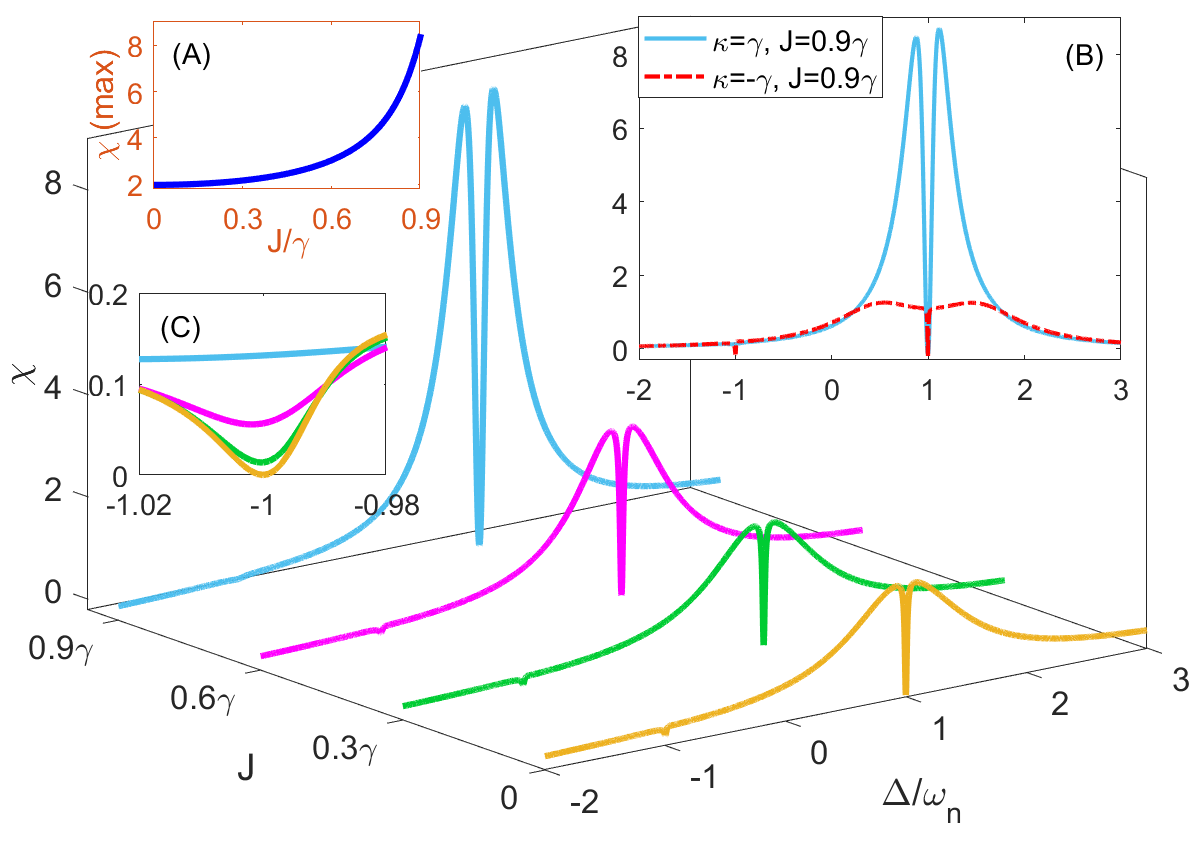} \centering
\caption{The absorption spectrum of the output field as a
function of $\Delta /\protect\omega _{n}$ and hoping factor J with $\protect%
\kappa =\protect\gamma$. The inset (A) represents the variation of maximum
value of absorption spectra with the increase of hoping factor. The inset
(B) represents the absorption profile when $\protect\kappa =\protect\gamma$
(solid blue line) and $\protect\kappa =-\protect\gamma$ (dot-dashed black
line). The inset (C) represents the absorption spectrum around $\Delta /%
\protect\omega _{n}=-1$. The others parameters are the same as in Fig. 2.}
\end{figure}
Next, we plot the absorption spectrum as a function of $\Delta/\omega_{n}$
and the photon hoping strength $J$. When $J=0$, there exists two paths for
the generation of photon at the probe frequency in the output field. One is
anti-Stokes scattering from the pump field and the other is the probe field
itself. The ONMS, in this situation, reduces to typical passive system and
there is transparency window at $\Delta/\omega_{n}=1$. However, when the
photon hoping strength is not zero, i.e., $J\neq0$, there exists an
additional rout in addition to two previous routs. The additional rout is
week probe field $\rightarrow$ passive cavity $\rightarrow$ active cavity $%
\rightarrow$ passive cavity $\rightarrow$ output. These three routs are
responsible for the enhancement of absorption spectra in the output. One can
see that only the strength of the absorption peaks (as shown by the inset
(A) of Fig. 4.) but also the width of the NMIT window gets broadened with the
strengthening $J$ as shown in Fig. 4. The inset (B) and (C), in Fig. 4.,
represent the absorption spectrum around $\Delta /\omega_{n}=-1$ and $\Delta
/\omega_{n}=1$, respectively. In addition, we also plot the absorption
profile when $\kappa =\gamma$ (solid blue line) and $\kappa =-\gamma$
(dot-dashed red line), in inset (B) of Fig. 4. It is worth noting that the
NMIT spectrum exists for both active-passive case and passive-passive case
as shown in the inset (B), in Fig. 4. In addition, the absorption peaks are
considerably higher for passive-passive compared to that of active-passive
ONMS.

\begin{figure*}[tbp]
\includegraphics[width=1.9\columnwidth,height=3.2in]{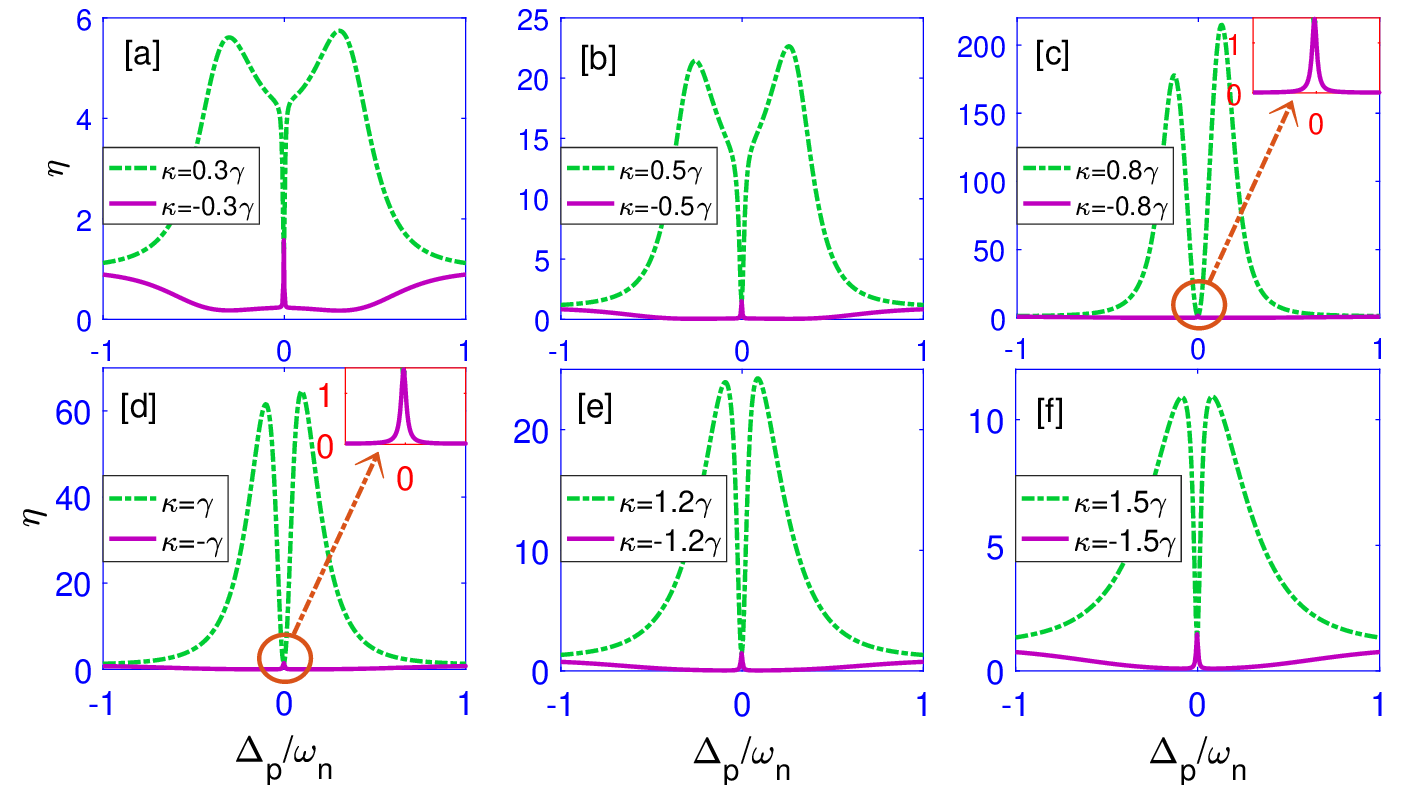} \centering
\caption{The transmission rate as a function of $\Delta /%
\protect\omega _{n}$ in the passive-passive (green dot-dashed line) and passive-active (solid purple line) system. The
others parameters are the same as in Fig. 2.}
\end{figure*}

\textbf{Gain dependence of the optical transmission}. For various values of the gain, we depict the transmission as a function of $\Delta_{p}/\omega_{n}$ in Figure 5. It is noteworthy that for $J=0.9\gamma$, the transmission at $\Delta /\omega_{n}=-1$ is very much small as compared to $\Delta /\omega_{n}=1$, therefore, we focus our discussion for the prob response around $\Delta /\omega_{n}=1$.
Fig. 5 illustrates how $\kappa/\gamma$ affects the optical transmission rate. In case of passive-active ONMS, as $\kappa/\gamma>0$ increases, the heights of the sideband peaks enhances until $\kappa/\gamma=0.8$, where maximum transmission at sidebands occurs. While in case of passive-passive, both sideband peaks are suppressed when the gain is increased further.
This is because, if $J>(\kappa+\gamma)/2$, the system remains in the $\mathcal{PT}$-symmetry phase; if $J<(\kappa+\gamma)/2$, on the other hand, the system transits to the $\mathcal{PT}$-symmetry-broken phase. Owing to the fact that in the $\mathcal{PT}$-symmetry phase, the criterion $J>\sqrt{\kappa\gamma}$ can always be satisfied. Nevertheless, if the system is in the $\mathcal{PT}$-symmetry-broken phase, $J>\sqrt{\kappa\gamma}$ is not always satisfied . Thus, by adjusting the gain in the active cavity, one can modulate the ONMS to transit from the NMIT to inverted NMIT. Thus, for $J/\gamma=0.9$, the ONMS stays in the $\mathcal{PT}$-symmetric phase for $\kappa/\gamma<0.8$, but for $\kappa/\gamma>0.8$, the ONMS shifts to the broken-$\mathcal{PT}$ phase.
Increasing the gain value above the phase transition value $\kappa/\gamma=0.8$ at $J=0.9\gamma$, the cavity field intensity in the passive cavity is notably reduced which weaken the strength of the ONMS interactions and therefore the value of the transmission. It has a striking resemblance to a previous experiment with two coupled resonators, which demonstrated that lowering (raising) the loss of one resonator below (above) a critical threshold decreases (increases) the other resonator's intracavity field strength, suppressing (enhancing) transmission \cite{BPNG}.
It's crucial to remember that increasing (decreasing) gain and decreasing (increasing) loss are
similar.
Hence, we conclude that the active NMIT in the $\mathcal{PT}$-regime can be considered as an analog of the optical inverted-EIT \cite{OTTM}.

\begin{figure}[t]
\includegraphics[width=1\columnwidth,height=1.7in]{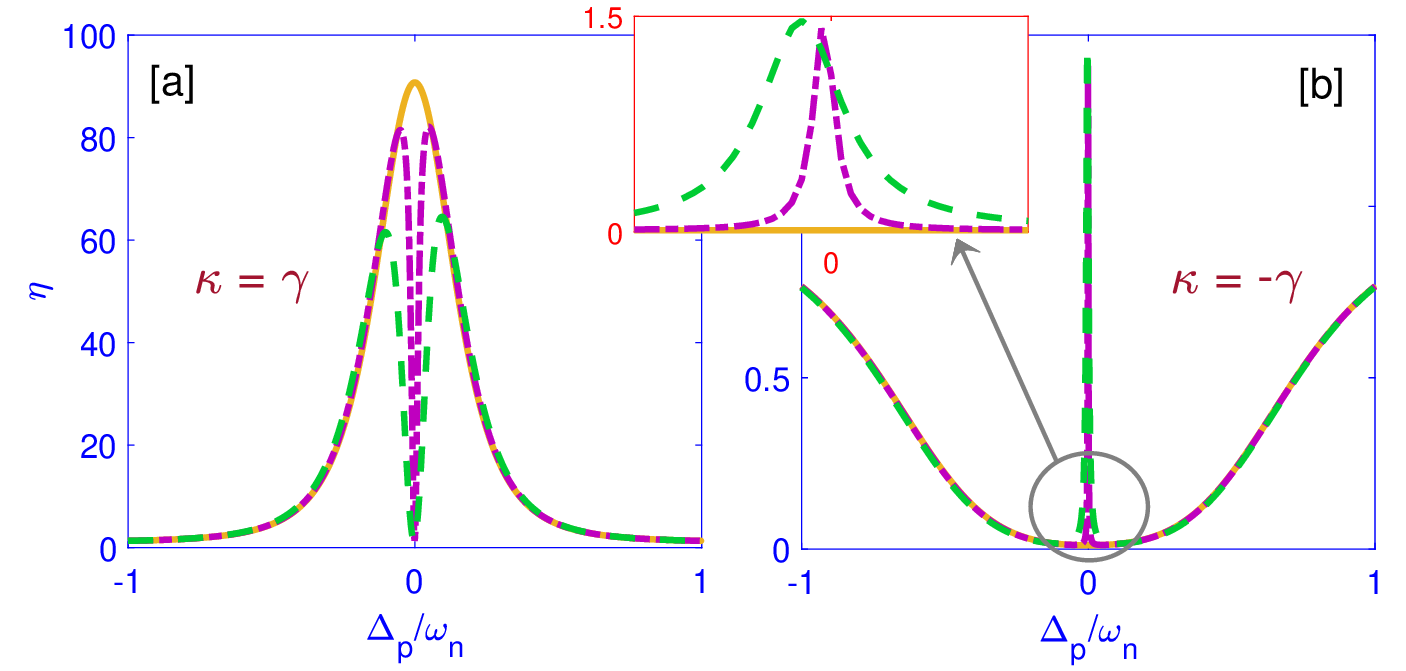} \centering
\caption{The transmission rate as a function of the
detuning, when sin$(2kx_{n})=0$ (solid yellow), sin$(2kx_{n})=0.05$ (dot-dashed
magenta) and sin$(2kx_{n})=0.1$ (dashed green)}
\end{figure}
\textbf{Opto-nanomechanical coupling dependence  of the optical transmission}. It is important to observe the effect of opto-nanomechanical coupling strength on the transmission rate. Since,  opto-nanomechanical coupling strength is the product of opto-nanomechanical coupling coefficient and the steady-state value of the cavity field, which depends upon the input laser field with amplitude $\Omega_{d}$, so
any variation of $\Omega_{d}$, through the power of input laser field or $G_{n}$ through the factor sin$\left( 2kx_{n}\right)$  directly effects the effective optonanocal coupling strength. Therefore, to investigate the effect of opto-nanomechanical coupling on the transmission rate, we have plotted the transmission rate as a function of $\Delta_{p}/\omega_{n}$ for different opto-nanomechanical coupling strengths by varying opto-nanomechanical coupling coefficient via sin$\left( 2kx_{n}\right)$.
For the
passive-active ONMS, increasing the opto-nanomechanical coupling leads to a notable
decrease of the sideband amplifications $\Delta_{p}/\omega_{n}=0$.
Furthermore, for the passive-passive ONMS, the width and the transmission rate around $\Delta/\omega_{n}=0$
increase with increasing opto-nanomechanical coupling (see Fig. 7(b)).
\section{CONCLUSIONS}
In this paper, we considered a $\mathcal{PT}$-symmetric two cavity system and studied the tunable optical response of the ONMS, containing a levitating nanosphere inside a passive cavity. ONMS system exhibits NMIT even in a situation with no loss/gain in the active cavity. We also observe an additional
absorption profile in this case. Furthermore, the absorption profile notably enhances as the hoping strength between the active and passive cavity increases. Next, for balanced gain/loss case (i.e., $\kappa/\gamma=1 $), height as well as the width of the NMIT window increase with increasing hoping factor $J$.
We found that the transmission spectrum depends on various system parameters, like, gain-to-loss ratio, detuning, hoping factor and the coupling between the nanosphere and right mirror of passive cavity.
An interesting feature is the optical response due to opto-nanomechanical coupling in $\mathcal{PT}$-symmetric and broken $\mathcal{PT}$-symmetric regimes has been discussed. Our results show that the presence of nanosphere in the passive-passive cavity enhances the width and transmission rate of NMIT window in passive-passive regime and in passive-active regime, a notable decrease of sideband amplification has been observed. In the end, we assert that this study may find some applications in controlling light propagation and optical switching.

\end{document}